# Does Publicity in the Science Press Drive Citations?

Manolis Antonoyiannakis[1,2]

[1] ma2529@columbia.edu

Department of Applied Physics & Applied Mathematics, Columbia University, 500 W. 120th St., Mudd 200, New York, NY 10027 (USA)

[2] American Physical Society, Editorial Office, 1 Research Road, Ridge, NY 11961 (USA)

**Introduction**

Publishing is a selection process. Through a series of selections, journal editors decide which papers merit external review, which papers to send back to referees, and which papers to publish. But increasingly over the past 20 years, the selection of papers does not stop at publication, as editors curate lists of their favorite accepted or recently published papers, sometimes accompanying them with short summaries or longer commentaries by experts.

The community is paying attention. Researchers often promote (in their websites or resumés) their highlighted papers, while funding agencies track their grantees' progress by monitoring coverage in highlighting platforms.

Citation metrics agree well with peer review at the aggregate level (Traag & Waltman, 2019). So, does the post-acceptance "round" of review of highlighted papers produce a citation advantage? We address this question quantitatively, for several publicity markers on papers published in the journal Physical Review Letters (PRL) of the American Physical Society (APS). We thus extend our previous work (Antonoyiannakis, 2015).

**Explanation of highlighting platforms**

*Cover image*: Chosen for aesthetic reasons mainly.
*Editors' Suggestions*: Chosen for potential interest.
*Viewpoints*: Commentaries commissioned by the APS Physics editors and written by experts. They explain why a paper is important to the field.
*Focus stories*: Journalist-written news stories that explain the latest research to non-physicists.
*Synopses*: Short summaries of newsworthy results written by journalists and APS Physics staff.
*Research Highlights*: Short summaries of papers written by journal editors in the *Nature* journals.
*News & Views*: Commentaries by experts or (less often) by journal editors in the *Nature* journals.

**Multiple Linear Regression**

We perform Multiple Linear Regression on the citation data for papers in these highlighting platforms. As a first sample, we chose papers published from 2008–2012 (Fig. 1). This amounts to 246 papers in the journal cover, 203 in Focus, 354 Viewpoints, 1232 Editors' Suggestions, 556 Synopses, 84 Research Highlights in Nature Physics, 36 News & Views in Nature Physics, and 91 Research Highlights in Nature. Clearly, the Viewpoint marker is the strongest predictor of citations. All else being equal, a Viewpoint marker allocates 10 additional citations to a PRL paper within 1 year of publication, and 44 citations within 5 years. The second strongest predictor of citations is Research Highlights in Nature. The third strongest predictors are Research Highlights in Nature Physics and Editors' Suggestions; and after that, Synopses and Covers. (The coefficients of News & Views in Nature Physics or Focus papers lack statistical significance.) We thus observe the following stratification of citation accrual:

$$\text{Viewpoint} > \text{RHNat} > \begin{cases} \text{Suggestion} \\ \text{RH NatPh} \\ \text{N\&V NatPh} \end{cases} > \begin{cases} \text{Synopsis} \\ \text{Cover} \\ \text{Focus} \end{cases} \quad (1)$$

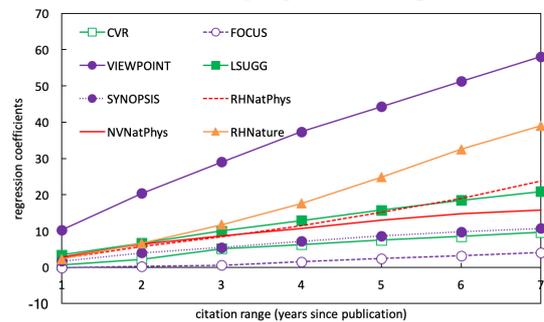

**Figure 1. Coefficients of multiple linear regression for each highlighting platform.**

Eq. (1) shows also a hierarchical pattern of decreasing scrutiny regarding importance during peer review. Viewpoints and Editors' Suggestions are internal highlights, for which the editors have the full benefit of peer review. Between these two, Viewpoints are vetted more, since they are discussed both among PRL journal editors and in a committee of Physics editors who seek further advice from experts on whether a Viewpoint is warranted. Synopses, Focus and Viewpoints are mutually exclusive, and since Viewpoints receive the highest level of scrutiny, Synopses and Focus articles are vetted less for importance. Research Highlights in Nature or Nature Physics, and News & Views in Nature Physics are all external highlights, which makes their selection more challenging, since their editors do not normally have access to the peer review files from PRL to aid their decisions. These journals are also more constrained in terms of space since their highlights cover several journals, and, for Nature, several fields. So, it makes sense that these platforms show a weaker prediction of citation accrual than Viewpoints, which receive the highest scrutiny among all platforms studied here. Finally, the difference in citation advantage between Research Highlights in Nature and Nature Physics may be because Nature is more selective than Nature Physics since it covers all fields of science.

The citation advantage of a Cover article, while statistically significant, is small. All else being equal, placement in the PRL cover adds no more than 1.5 additional citations per year. Thus, accidental or serendipitous publicity (recall that covers are chosen for aesthetics mainly) brings a small citation advantage, in direct analogy to the effect of visibility bias reported by Ginsparg and Haque (2009), for papers that accidentally end up in the top slot of the daily arXiv email listings. However, the citation advantage is clearly greater when publicity is deliberate and results from an endorsement of the paper's merit formed through peer review, as in a Viewpoint. So, publicity alone does help in terms of citations, but does not make a big difference unless it is supported by an endorsement, formed through peer review, that the paper has above-average merit.

**Citation medians**

We can also compare medians. In Fig. 2 we show the annual median citation advantage for each highlighting platform. These results confirm our findings from multiple linear regression analysis.

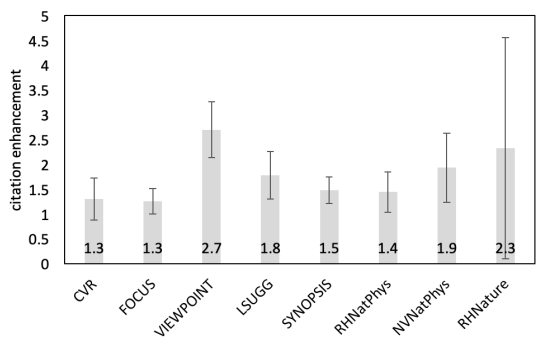

**Figure 2. Annual enhancement of median citations per highlighting platform compared to the PRL journal. Publications from 2008–2018. Citation range 1–9 years after publication. Error bars at 99% confidence interval.**

**Clarivate list of highly cited papers**

Can we use highlighting markers as predictors of a paper being *highly* cited? This question takes us beyond the previous analysis, since placement of a paper in a highly cited list is identifying extreme, not average, citation performance. As a benchmark for highly cited papers, we use the Highly Cited Papers (HCP) indicator of Clarivate Analytics. These papers are at the top 1% cited in their subject per year. We downloaded the list of HCP papers in physics published from 2010–2018, among which there are 1371 PRL papers. In Table 1 we show how well each platform predicts placement in the HCP list, i.e., how often highlighted papers are highly cited, which is the positive predictive value, or precision. The hierarchical pattern of Eq. (1) is thus reproduced. Again, we find that highlighting for importance correlates with citations—even for the extreme case of top-1% cited papers.

**Table 1. Summary statistics for the precision (positive predictive value) of each platform in identifying highly cited papers.**

| Highlight | Count | Precision |
|---|---|---|
| CVR | 446 | 0.099 |
| FOCUS | 378 | 0.063 |
| VIEWPOINT | 638 | 0.290 |
| LSUGG | 3269 | 0.143 |
| SYNOPSIS | 1117 | 0.117 |
| RHNatPhys | 116 | 0.164 |
| NVNatPhys | 33 | 0.121 |
| RHNature | 109 | 0.239 |

**Conclusions**

Our key conclusion is twofold. First, highlighting for importance identifies a citation advantage. Accidental or serendipitous publicity (i.e., mere visibility), gives a clearly smaller citation advantage. Second, the stratification of citations for highlighted papers follows the degree of vetting for importance during peer review. This implies that we can view the various highlighting platforms as predictors of citation accrual, with varying degrees of strength that reflect each platform's vetting level. More details are provided in Antonoyiannakis, 2021.